\documentclass[aps,prb,twocolumn,longbibliography,groupedaddress,showpacs,showkeys,citeautoscript,amsfonts,amsmath,amssymb,flushbottom,floatfix,10pt]{revtex4-1}

\usepackage{graphicx}
\usepackage{nicefrac}
\usepackage[utf8]{inputenc}
\usepackage[T1]{fontenc}
\usepackage{mathrsfs}
\usepackage{wasysym}
\usepackage{url}

\newif\ifshowcolor

\ifshowcolor

\else

\fi

\begin{document}
\setlength{\floatsep}{0pt}
\setlength{\intextsep}{0pt}

\newcommand{\scri}{{\mathscr I}}

\newcommand{\bskipdm}{\mskip -3.8\thickmuskip}
\newcommand{\bskiptm}{\mskip -3.0\thickmuskip}
\newcommand{\bskipsm}{\mskip -3.1\thickmuskip}
\newcommand{\bskipssm}{\mskip -3.1\thickmuskip}
\newcommand{\fskipdm}{\mskip 3.8\thickmuskip}
\newcommand{\fskiptm}{\mskip 3.0\thickmuskip}
\newcommand{\fskipsm}{\mskip 3.1\thickmuskip}
\newcommand{\fskipssm}{\mskip 3.1\thickmuskip}
\newcommand{\pint}{\mathop{\mathchoice{-\bskipdm\int}{-\bskiptm\int}{-\bskipsm\int}{-\bskipssm\int}}}
\newcommand{\ds}{\displaystyle}
\newcommand{\D}{\mathrm{d}}
\newcommand{\I}{\mathrm{i}}
\newcommand{\EXP}[1]{\mathrm{e}^{#1}}    

\newcommand{\tpp}{\hat{t}}
\newcommand{\rpp}{\hat{r}}
\newcommand{\zpp}{\hat{z}}
\newcommand{\ppp}{\hat{\varphi}}
\newcommand{\tbb}{\bar{t}}
\newcommand{\rbb}{\bar{r}}
\newcommand{\zbb}{\bar{z}}
\newcommand{\pbb}{\bar{\varphi}}
\newcommand{\lbb}{\bar{\ell}}
\newcommand{\ttt}{\tilde{t}}
\newcommand{\rtt}{\tilde{r}}
\newcommand{\ztt}{\tilde{z}}
\newcommand{\ptt}{\tilde{\varphi}}
\newcommand{\abs}[1]{\left\lvert #1 \right\rvert}
\newcommand{\tvv}{\check{t}}
\newcommand{\xvv}{\check{x}}
\newcommand{\omvv}{\check{\omega}}
\newcommand{\kvv}{\check{k}}
\newcommand{\vvv}{\check{v}}
\newcommand{\lvv}{\check{\ell}}

\newcommand{\abl}[3][\empty]{\frac{\D^{#1} #2}{\D {#3}^{#1}}}
\newcommand{\pabl}[3][\empty]{\frac{\partial^{#1} #2}{\partial {#3}^{#1}}}

\title{The Vaidya metric: expected and unexpected traits of evaporating black holes}

\author{Julius Piesnack and Klaus Kassner
}
\affiliation{Institut für Physik,  Otto-von-Guericke-Universität, Magdeburg, Germany
}

\begin{abstract}
  The ingoing Vaidya metric is introduced as a model for a
  non-rotating uncharged black hole emitting Hawking radiation. This
  metric is expected to capture the physics of the spacetime for
  radial coordinates up to a small multiple $(>1)$ of the
  Schwarzschild radius. For larger radii, it will give an excellent
  approximation to the spacetime geometry in the case of astrophysical
  black holes $(M\ge M_{\astrosun})$, except at extremely large
  distances from the horizon (exceeding the cosmic particle horizon).
  In the classroom, the model may serve as a first exploration of
  non-stationary gravitational fields.  Several interesting
  predictions are developed. First, particles dropped early enough
  before complete evaporation of the black hole cross its horizon as
  easily as with an eternal black hole. Second, the Schwarzschild
  radius takes on the properties of an apparent horizon, and the true
  event horizon of the black hole is \emph{inside} of it, because
  light can escape from the shrinking apparent horizon. Third, a
  particle released from rest close enough to the apparent horizon is
  strongly repelled and may escape to infinity. An interpretation is
  given, demonstrating that such a particle would be able to compete,
  for a short time, in a race with a photon.
\end{abstract}

\date{August 30, 2021}

\keywords{Black holes, Hawking radiation, evolving horizons}

\maketitle

\section{Introduction}

Schwarzschild black holes\cite{schwarzschild16a,droste17} are among
the most popular examples of spacetimes studied in general relativity
courses. Their metric is simple, enables a straightforward discussion
of the four classical tests of general relativity, and allows one to
explore some strong-field effects as well as the concept of an event
horizon.

The metric is stationary, so effects of \emph{time-dependent}
gravitational fields are absent. Beyond that, the interesting
phenomenon of gravitational \emph{repulsion} of radially infalling
particles does not arise in the Schwarzschild metric. It happens for
\emph{charged} spherically symmetric black holes, described by the
Reissner-Nordström metric \cite{gron86}, but only well inside the
event horizon. Some non-stationary metrics turn out to exhibit it in
more accessible places.

A simple time-dependent metric is the Vaidya metric,\cite{vaidya51}
describing the exterior of a spherically symmetric mass distribution
that absorbs or emits null dust, i.e., massless particles.  While
spectacular time-dependent effects may be discussed in studying
the properties of gravitational waves,\cite{einstein37} the Vaidya
metric is not devoid of some interesting ones either.

Note that the Vaidya metric is not purely of academic interest. It
arises in a natural way in phenomenological models of a black hole
evaporating by Hawking radiation, an object that is not fully
describable by a stationary metric. Since Vaidya metrics are models of
the \emph{spacetime} resulting from Hawking radiation, not of the radiation
itself (which is mostly electromagnetic in nature), Hawking radiation
will be hardly touched upon in the main part of this article. Rather,
we will give a discussion of its essential features in
Sec.~\ref{sec:hawking_radiation}, preceding the presentation of our
model and its results.

In 2005, Aste and Trautmann considered a toy model,\cite{aste05} in
which they made the mass term appearing in the Schwarzschild metric
dependent on the global time. With this metric, they found that the
black hole evaporates under an infalling particle, before the latter
can cross the horizon.  It was pointed out\cite{kassner19} that this
model is not physically meaningful; the Schwarzschild time coordinate
inside the event horizon is unrelated to the time coordinate outside
due to the fact that it is not continuous across the horizon
(diverging to infinity there).

An improvement was suggested in Ref.~\onlinecite{kassner19}, where
Gullstrand-Painlevé (GP) coordinates \cite{painleve21,gullstrand22}
were taken as a starting point to introduce a time-dependent mass
parameter. These coordinates are continuous across the event horizon
of the Schwarzschild spacetime, so making the mass time-dependent at
least has a well-defined meaning. In this model, an observer falling
towards an evaporating black hole will ordinarily cross the
Schwarz\-schild radius without any problem (except close to the time
of final dissolution of the hole) and then hit the central singularity
in finite proper time. Clearly, this is still a toy model with a
number of unrealistic features. For example, since the mass parameter
depends on the time coordinate only, it will change instantaneously
for all observers sitting at different radial positions.  However,
energy is radiated away at a finite velocity. Therefore, two
coordinate stationary observers with different radii should
experience, at the same time, the attraction of different masses
inside their respective radial shells.

This problem may be solved by using Eddington-Finkelstein (EF)
coordinates\cite{eddington24,finkelstein58} instead. Making the mass
dependent on the EF time leads to a Vaidya metric,
in which the mass parameter is constant not on a spacelike but on a
null surface. Because this means that the radiated energy moves at the
speed of light, compatibility with causality is achieved, and the
emitted particles must be massless, such as photons. The construction
of such a metric is still a far cry from producing a dynamical
spacetime that would describe an evaporating black hole in a quantum
mechanically consistent way.  We do not know the time dependence of
Hawking radiation at times close to the -- presumed -- final
disappearance of the black hole, where quantum effects should become
strong, and we cannot calculate the back reaction between the Hawking
radiation flux and the evolving metric non-perturbatively. Even a
semiclassical self-consistent solution does not seem to be available,
in which the metric would be a classical field arising from the
quantum mechanical expectation value
$\left\langle T_{\mu\nu}\right\rangle$ of the stress-energy
tensor.\cite{hiscock81a} Nevertheless, the simplest guess at what the
semiclassical time-dependent metric of an evaporating black hole
should look like, subject to the condition of a minimum of realism
regarding energy transport, would be a Vaidya metric, leaving one with
the freedom to specify largely arbitrary time-dependent mass
functions.\cite{hiscock81a}

As will be discussed below, if the dynamical horizon of the
description is to correspond to a future horizon, the metric must be
an ingoing Vaidya metric. Such a metric with a mass parameter that
decreases as a function of time corresponds to a black hole losing
mass by infalling negative-energy null dust. This feature, while
acceptable near the horizon (as will be seen in the next section), is
not very realistic far from it, where Hawking radiation clearly is
outgoing and consists of positive-energy particles.  Then, to add more
realism to the model, the ingoing Vaidya metric near the horizon
should be complemented by an outgoing one at larger radial
coordinates, with the surface joining the two metrics corresponding to
a pair creation location.\cite{hiscock81b} A few years ago, a model
for an evaporating black hole without any spacetime singularities was
presented that employs these features.\cite{hayward06} Since our
purpose is to present a model that is useful in exploring physical
effects in the classroom, we stick to the simplest case discussed by
Hiscock\cite{hiscock81a} containing some elements of realism, i.e.,
just an ingoing Vaidya metric. Local results near the horizon then
should have validity, but the assessment of results obtained at large
radii needs additional scrutiny.

The remainder of the paper is organized as follows. After the
explanatory Sec.~\ref{sec:hawking_radiation} on Hawking radiation, we
consider, in Sec.~\ref{sec:schwarzschild_eddington_finkelst}, the
maximally extended Schwarzschild metric in a diagram using
Kruskal-Sze\-keres (KS) coordinates\cite{kruskal60,szekeres60}. Discussing
which regions of the full spacetime are covered by ingoing and
outgoing EF coordinates, we decide which of the
two sets is the best starting point in constructing a Vaidya metric.
Section \ref{sec:vaidya_eq_motion} gives the equations of motion for a
test particle falling radially into the model black hole and
simplifies them to a form that can be compared with the Newtonian
limit. These equations are solved numerically in
Sec.~\ref{sec:num_sol_discuss}.  
The existence of repulsive effects
of the time-dependent gravitational field is demonstrated
analytically. Finally, some conclusions are given in
Sec.~\ref{sec:conclus}.

\section{Hawking radiation}
\label{sec:hawking_radiation}

Hawking provided a physical mechanism for Bekenstein's idea that the
area of the event horizon of a black hole corresponds to its entropy
and its surface gravity to its
temperature\cite{bekenstein73,bekenstein74}
(each with the appropriate proportionality constant).
In two papers exploring this quantum
mechanical mechanism, \cite{hawking74,hawking75} he presented a
calculation based on quantum field theoretical considerations,
according to which a field operator associated with a vacuum state in
Minkowski spacetime (at past null infinity $\scri^{-}$) will, after
the spacetime has become curved due to the formation of a black hole,
be associated with a state of non-zero particle content near the event
horizon. Some of these particles, mostly photons, escape to future
null infinity $\scri^{+}$, and the calculation proceeds by following
the field at $\scri^{+}$ backwards in time through the interaction
with the collapsing star to $\scri^{-}$, using analytic continuation
techniques to evaluate the spectral properties of the so-created
radiation, which turns out to be thermal with temperature and entropy
given by the Bekenstein expressions. The field calculation may be
visualized in terms of a wave picture, with the properties of the
ingoing wave coming from $\scri^{-}$ modified by its interaction with
the event horizon. Detailed considerations involve negative frequency
components of the wave falling into the black hole and positive
frequency components being amplified%
, allowing part of them to escape
to~$\scri^{+}$.

It should be noted that Hawking himself provided an alternative
picture of the process already in the extended version of his first
explanation of the effect.\cite{hawking75}
 It focuses on the particle
content of the quantum fields involved rather than on the
decomposition of the field into waves with amplitudes given by
annihilation and creation operators. In this picture, Hawking described a
negative energy flux inward across the horizon, coming from the
surrounding region in which, by virtue of quantum fluctuations, pairs
of virtual particles appear and disappear. One of these typically has
positive and the other negative energy (so as to satisfy energy
conservation in the long run). He then stated:\cite{hawking75}
\emph{The negative particle is in a region which is classically
  forbidden but it can tunnel through the event horizon to the region
  inside the black hole where the Killing vector which represents time
  translations is spacelike. In this region the particle can exist as
  a real particle with a timelike momentum vector even though its
  energy relative to infinity as measured by the time translation
  Killing vector is negative. The other particle of the pair, having a
  positive energy, can escape to infinity where it constitutes a part
  of the thermal emission described above.
  [...]
  Instead of
  thinking of negative energy particles tunnelling through the horizon
  in the positive sense of time one could regard them as positive
  energy particles crossing the horizon on past-directed world-lines
  and then being scattered on to future-directed world-lines by the
  gravitational field.} Although Hawking emphasized that this is a
heuristic picture, not to be taken too literally, it may be more
appealing than the picture of positive-frequency and negative-frequency wave
components, where the energy balance is not readily visible with the
overall wave consisting predominantly of positive-frequency
components.

Of course, when talking of particles here, we mean \emph{quantum}
particles without committing ourselves to their manifestation in a
localized or wavy manner. In fact, the particles of Hawking radiation
are not very pointlike.  The wavelength of the maximum of the spectrum
of Hawking radiation for a solar-mass black hole is about 47~km at
infinity, whereas the Schwarzschild radius is a mere 3 km. So these
photons would be detectable with antennas rather than with
photomultipliers, i.e., they would be considered waves rather than
particles. Moreover, they are so poorly localized that it is 
impossible to indicate a single value for the redshift they have
undergone since their creation near the black hole horizon (because
the redshift is significantly different for different parts of the
wave). A wave emerging at about two Schwarzschild radii from the
center of the geometry will have only a roughly 40\% higher frequency
at that position than at infinity, meaning that its most probable
wavelength would still be more than 30 km. Photons with such
wavelengths might easily tunnel distances on the order of 10 km, i.e.,
three Schwarzschild radii. Hence, in adopting the particle picture, we
should not imagine the radiation to come from the surface constituted
by the horizon. Its location of creation may be washed out with an
uncertainty on the order of the Schwarzschild radius. In fact, the
``quantum atmosphere'', in which Hawking radiation is created, has
recently been estimated, from a $1+1$ dimensional calculation, to
extend in radial coordinate from 1.5 to 2 Schwarzschild
radii.\cite{dey19}

Newer calculations of Hawking radiation have
been presented\cite{parikh00} which are much closer in spirit to the
picture of tunneling quantum particles than Hawking's original one.
One advantage of calculations that are manifestly based on a tunneling
process is that it is not necessary to consider the complete collapse
geometry before formation of the black hole. While the details of the
collapse do not influence the final result, the picture usually drawn,
having null trajectories in a Penrose diagram that connect $\scri^{-}$
with $\scri^{+}$ (and pile up near the horizon) may favor
misconceptions such as the idea that all of the Hawking radiation
comes from the time before the surface of the collapsing star crosses
the horizon,\cite{gerlach76} which would mean that the black hole
remains \emph{incipient} forever, i.e., that it does not actually form. 
A nice compact discussion of features of Hawking radiation may
be found in Ref.~\onlinecite{visser03}.

Experimental verification of Hawking radiation from a typical black hole
with a mass exceeding that of our sun is virtually impossible,
because the effect is so tiny. It has been suggested\cite{unruh81}
that \emph{sonic} black holes (in which a fluid takes on supersonic
speeds so that there should be Hawking-like sound emission from the
``sonic horizon'') might allow experimental access to the
phenomenon. Indeed, success in detecting Hawking radiation from a
Bose-Einstein condensate as a black-hole analog was reported
recently.\cite{steinhauer16}

Returning to real black holes, an object falling towards the event horizon
seems to get slower and freeze there due to the fact that photons take
longer and longer to escape from the vicinity of the horizon.
What is more, it turns out that the object will take \emph{infinite}
\emph{Schwarzschild time} to actually reach the horizon, so the
slowing-down would, it appears, not just be an optical illusion.
That a black hole dissolves via Hawking radiation in \emph{finite}
time then leads to an apparent paradox: If it takes infinite time for
an observer to fall into an eternal black hole with fixed event
horizon, falling into an evaporating black hole should also take
infinite time. After all, the attracting mass decreases and the
horizon recedes, so if anything, one should expect an infaller to
take longer for the evaporating black hole than for the static one.
But then the black hole must disappear, \emph{before} the observer can
fall into it. This of course begs the question how an event horizon
could form in the first place, because the surface of the star will,
on collapse, behave similarly to the infaller. On the other hand,
observers can fall into eternal black holes within a finite interval
of their proper time and Hawking radiation is a weak effect for
stellar-mass black holes, which suggests that crossing of the event
horizon by an observer should happen essentially in the same way as
for a static black hole. But it is a contradiction for an observer to
both fall and not fall into a black hole. This paradox has been
resolved in Ref.~\onlinecite{kassner19} modeling the metric of an
evaporating black hole appropriately, by use of a time coordinate that
does not diverge at the horizon and is resolved in this paper the same
way (but using a different time coordinate).

In such a model, there must be a region of negative energy density
outside the horizon, as transpires from the quote reproduced above
from Hawking's 1975 paper.\cite{hawking75} In classical general
relativity, negative energy density is considered forbidden and
solutions requiring
it 
are viewed as unrealizable, but in the presence of quantum mechanics,
they cannot be discarded automatically. Our considerations in the
following will be based on a classical model, but negative energy
density is admitted in order to mimic the quantum effect of
negative-energy virtual particles. 

\section{The Schwarzschild spacetime in terms of Eddington-Finkelstein coordinates}
\label{sec:schwarzschild_eddington_finkelst}
In this section, we illustrate the idea that different coordinate
systems may cover different parts of the spacetime manifold.
The necessity to capture certain features (such as a future event
horizon) in the description may then restrict the choice of possible
coordinates.

\begin{figure}[thb]
  \vspace{3mm}
  \includegraphics[width=8cm]{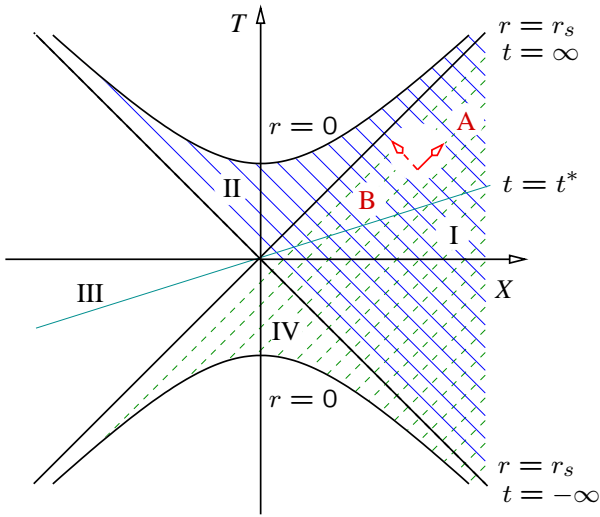}
  \caption{Kruskal-Szekeres diagram of the maximally extended
    Schwarzschild metric.  Loci of constant Schwarzschild time
    coordinate $t$ correspond to straight lines through the origin (a
    line $t=t^*$ is drawn explicitly). Loci of constant Schwarzschild
    radial coordinate $r$ are given by hyperbolas having the angle
    bisectors of the axes as asymptotes (the lines $r=0$ are drawn)
    or, in the limit $r=r_s$, by these bisectors themselves. Two sets
    of Schwarzschild coordinates are necessary to cover all four
    regions. The solid hatch lines correspond to (non-equidistant)
    constant null coordinates $v$ in the ingoing Eddington-Finkelstein
    metric, the dashed ones to constant coordinates $u$ in the
    outgoing EF metric. Other features are
    explained in the text.}
\vspace{2mm}
  \label{fig:KS_diagram}
\end{figure}

The maximally extended Schwarzschild solution, corresponding to an
eternal spherically symmetric black hole -- and a bit more -- is
describable via KS coordinates.  The
resulting diagram given in Fig.~\ref{fig:KS_diagram} will also be useful in exhibiting the
difference between the two EF coordinate
systems.

The metric is\cite{mueller10}
\begin{align}
  \D s^2 = \frac{4 r_s^3}{r} \,\EXP{-r/r_s} \left(\D T^2-\D X^2\right)- r^2 \D\Omega^2\>,
  \label{eq:KS_metric}
\end{align}
where $r_s=2GM/c^2$ defines the Schwarzschild radius ($M$ is the black
hole mass, $G$ Newton's gravitational constant, $c$ the vacuum speed
of light) and $\D\Omega^2=\D\vartheta^2 +\sin^2\vartheta \D\varphi^2$
abbreviates the line element on a unit sphere. The radial coordinate
$r$ of the Schwarzschild metric is expressible via $X$ and $T$ with
the help of $X^2-T^2 = \left(\frac{r}{r_s}-1\right)\EXP{r/r_s}$.

The diagram of Fig.~\ref{fig:KS_diagram} may be read as giving a
two-dimensional section of the spacetime at fixed $\vartheta$ and
$\varphi$.  One of its interesting features is that radial light
rays are parallel to either of the two bisectors of the pair of
coordinate axes, so it is easy to discuss the exchange of light
signals of observers in the geometry.\cite{kassner17c} 
Event horizons are given in Fig.~1 by the future and past light
cones of the origin. The Schwarzschild time coordinate is timelike
only in the regions I and III ($X>\abs{T}$ and $X<-\abs{T}$,
respectively) but spacelike in the regions II and IV ($T>\abs{X}$ and
$T<-\abs{X}$, respectively, with $T^2-X^2<1$), because the absolute
value of the slope of any straight line through the origin is bigger
than 1 in these two regions.\footnote{A coordinate is timelike
  (spacelike) if a hypersurface of constant value of that coordinate
  is spacelike (timelike).} Correspondingly, the coordinate $r$
behaves like a spatial coordinate in regions I and III, but like a
time coordinate in regions II and IV.

Customarily, region I ($r>r_s$, $-\infty<t<\infty$) is considered as
describing the outside of a Schwarzschild black hole. The complete set
of Schwarzschild coordinates, including $r<r_s$, $-\infty<t<\infty$,
then either describes regions I and II or regions I and IV. That
Schwarzschild coordinates become singular at $r=r_s$ may be seen from
the fact that all points with $r=r_s$ (except $(X,T)=(0,0)$) also have
either $t=\infty$ or $t=-\infty$, so the pair $(r,t)=(r_s,\infty)$
actually describes an infinite set of events instead of a single one
(and so does $(r,t)=(r_s,-\infty)$).\footnote{The single event
  $(X,T)=(0,0)$ has an infinity of time coordinates $t$, just as the
  pole of a sphere has an infinity of $\varphi$ values in spherical
  coordinates. \label{foot:indef_time_coord}} The patches of
Schwarzschild coordinates separated by $r=r_s$ are not continuously
connected ($t$ becomes infinite between them), therefore the ambiguity
whether $r<r_s$ corresponds to region II or IV. In practice,
identification is always made with region II, because the horizon
between regions IV and I, sometimes called
antihorizon,\cite{hamilton18} is permeable from IV to I only and thus
corresponds to a \emph{white hole} (into which nothing can fall, but
from which stuff may be ejected). 
The vacuum solution, described by the metric \eqref{eq:KS_metric} is
not applicable inside a star, so observers in region I will see an
\emph{illusory horizon}\cite{hamilton18} (the surface of the
collapsing star) instead of an antihorizon.  Once a black hole has
formed, the vacuum part of the metric will resemble that of an eternal
black hole with only regions I and II explorable.

A pair of photons, one escaping one infalling, is indicated in the figure,
the first drawn as a solid arrow; the second as a dashed arrow.\footnote{The
  representation is symbolical, because pair creation should, as noted
  before, not be considered a pointlike event.} This is to illustrate
the point that for an observer $A$, whose approximate position is
marked in the figure, Hawking radiation seems to come from the
illusory horizon,\cite{hamilton18} because he will not see the
infalling photon and the trajectory of the escaping one is
\emph{parallel} to the horizon in the KS diagram. However, an observer
$B$ who is closer to the illusory horizon will not see that particular
photon, because it does not really come from the surface of the
collapsing star. For a shrinking black hole, observer $A$ will see
radiation that was emitted at a slightly higher temperature than that
seen by $B$,\footnote{However, since $A$ is at a larger $r$ coordinate
  than $B$, the radiation he sees will be more strongly redshifted
  than that seen by $B$.} and the difference in photon content was
created in the spacetime interval between the two observers, which
typically will mean that it was created after the observation made at
$B$'s spacetime position but before the observation by $A$.
An observer falling freely through the horizon should
not detect any Hawking radiation at all at the crossing point, if the
equivalence principle continues to hold at the horizon, as suggested
by Ref.~\onlinecite{singleton11}.\footnote{A freely falling observer
  is in a local inertial system, so local experiments do not allow her
  to detect the black hole horizon on crossing it. If the horizon is a
  condition for Hawking radiation in an almost stationary spacetime,
  then a freely falling observer cannot see radiation from that
  horizon, once she reaches it. Radiation from a distant horizon is
  not forbidden by the equivalence principle, as demonstrated by the
  simple fact that an observer at infinity does observe Hawking
  radiation. Ref.~\protect\onlinecite{singleton11} gives support to
  the validity of the equivalence principle at the horizon only for
  certain accelerated observers, by showing that Hawking radiation and
  Unruh radiation have the same temperature for them. Note that the
  calculation of a non-zero Hawking
  temperature\protect\cite{hamilton18} for a freely falling observer
  passing the horizon is insufficient to prove observability of the
  radiation. What has to be proven in addition is that its intensity
  (or its gray-body factor) is non-zero.}

Let us now discuss EF coordinates. One way to obtain them from
Schwarzschild coordinates is to first introduce the \emph{tortoise
  coordinate}
\begin{align}
  r^* = r+r_s \ln\abs{\frac{r}{r_s}-1}\>,
  \label{eq:r_tortoise}
\end{align}
which approaches $-\infty$ as $r\to r_s$, and then to introduce, as
new time coordinate, either
\begin{align}
  v= t+\frac{r^*}{c}\>,
  \label{eq:advanced_time}
\end{align}
yielding the ingoing EF metric with coordinates $v$, $r$, $\vartheta$,
and $\varphi$ or
\begin{align}
  u= t-\frac{r^*}{c}\>,
   \label{eq:retarded_time}
\end{align}
which gives the outgoing EF metric with coordinates $u$, $r$,
$\vartheta$, and $\varphi$.  It is then easy to determine the
coordinate transformations leading from the KS metric to the EF
metrics. These are
{\allowdisplaybreaks\begin{align}
  X &= \frac12 \EXP{cv/2 r_s} + \frac12 \left(\frac{r}{r_s}-1\right) \EXP{r/r_s-cv/2 r_s}\>,
      \nonumber\\
  T &= \frac12 \EXP{cv/2 r_s} - \frac12 \left(\frac{r}{r_s}-1\right) \EXP{r/r_s-cv/2 r_s}\>,
      \label{eq:EF_ingoing_KS}
\end{align}}%
and
{\allowdisplaybreaks\begin{align}
  X &= \frac12 \EXP{-cu/2 r_s} + \frac12 \left(\frac{r}{r_s}-1\right) \EXP{r/r_s+cu/2 r_s}\>,
      \nonumber\\
  T &= -\frac12 \EXP{-cu/2 r_s} + \frac12 \left(\frac{r}{r_s}-1\right) \EXP{r/r_s+cu/2 r_s}\>,
      \label{eq:EF_outgoing_KS}
\end{align}}%
respectively.  We note that all values $-\infty<v<\infty$ and
$0<r<\infty$ are admissible in the transformation
\eqref{eq:EF_ingoing_KS}. Moreover, for $r>r_s$, we have $X>0$, which
excludes region III, and for $r<r_s$, we have $T>0$, which excludes
region IV. Therefore, ingoing EF coordinates cover regions I and II
continuously. This is depicted in Fig.~1 by the region with solid
hatch lines.  Since constant $v$ implies that $X+T=\text{const.}$ (as
follows immediately from \eqref{eq:EF_ingoing_KS}), the hatch lines
also correspond to lines of constant coordinate $v$. So $v$ is a null
coordinate rather than a timelike one; at a fixed position $r$, it may
nonetheless nicely serve as a time coordinate.

Similar considerations for equations
\eqref{eq:EF_outgoing_KS} show that 
outgoing EF coordinates cover regions I and IV continuously; they
describe a white-hole metric. The dashed hatching depicts their region
of validity 
and dashed lines correspond to lines of constant coordinate~$u$.

It is then clear that if we wish to describe an evaporating
\emph{black} hole (with a future horizon), our starting point should
not be outgoing but ingoing EF coordinates. In terms of these, the
Schwarzschild line element reads
\begin{align}
  \D s^2 = \left(1-\frac{r_s}{r}\right) c^2 \D v^2 - 2 c \D v \D r - r^2 \D\Omega^2\>.
  \label{eq:EF_metric}
\end{align}
This metric is nonsingular at $r=r_s$, in spite of the fact that the
prefactor of $ \D v^2$ vanishes there. The non-diagonal term
$\propto \D v \D r$ ensures that none of the eigenvalues of the metric
become zero at $r=r_s$.

\section{Ingoing Vaidya metric and equations of motion}
\label{sec:vaidya_eq_motion}

A model for an evaporating black hole is obtained by
allowing the Schwarzschild radius to become $v$-de\-pend\-ent, which
produces an ingoing Vaidya metric:
\begin{align}
  \D s^2 = \left(1-\frac{r_s(v)}{r}\right) c^2 \D v^2 - 2 c \D v \D r - r^2 \D\Omega^2\>.
  \label{eq:vaidya_metric}
\end{align}
Our goal is then to study the equations of motions of particles moving
in this metric, in order to become conversant with the geodesic motion
in this non-stationary setting.

The Vaidya metric is not a vacuum solution. However, only one of the
covariant components of the Ricci tensor is non-zero:
$R_{vv}=r'_s(v) c/r^2 = 2 G M'(v)/r^2 c$. The scalar curvature
vanishes. Hence, the stress-energy tensor also has only one
nonvanishing element in the coordinates used in \eqref{eq:vaidya_metric},
which is given by $T_{vv}=M'(v) c^3/4\pi r^2$. Note that this is
negative, if the mass $M$ is a decreasing function of $v$.

Consider the line of constant Schwarzschild time $t=t^*$ drawn in
Fig.~\ref{fig:KS_diagram}. Moving upward to the right on this line
means increasing $v$, as $t$ remains constant but $r$ increases
(Eq.~\eqref{eq:advanced_time}).  This means that observers at
larger $r$ and fixed $t=t^*$ see smaller masses inside a sphere of
radius $r$, as $M(v)$ diminishes with increasing $v$. This must be so
for the ingoing Vaidya metric, because with increasing $r$, more and
more negative-energy null dust is between the observer at $r$ and the
horizon.

However, this is not what is to be expected far from the horizon for a
true evaporating black hole. Hawking radiation is outgoing and has
positive energy density, so the mass inside a sphere of radius $r$
should increase with $r$ at constant $t$. Note that this would be the
case if the mass depended on the coordinate $u$ of outgoing EF
coordinates. The value of $u$ is infinite at $X=T$ (see
Eq.~\eqref{eq:EF_outgoing_KS}) and decreases along the line $t=t^*$ as
one moves to the right (Eq.~\eqref{eq:retarded_time}). Therefore, far
from the horizon, the outgoing Vaidya metric is more compatible with
Hawking radiation than the ingoing one. But near the horizon, the
retarded Vaidya time $u$ becomes singular, so for a description
including the horizon and its interior, it is not an option, whereas
the ingoing Vaidya metric works fine and its negative energy density agrees
with quantum mechanical considerations. We do not switch between the
two metrics (at some prescribed surface) as was done in
Refs.~\onlinecite{hiscock81b,hayward06}, because the model would
become unnecessarily complicated. The effects to be discussed here
will be (slightly) modified quantitatively by such a model improvement
but not qualitatively. A detailed discussion of the quality of the
approximation is given after Eq.~\eqref{eq:ddot_r}, at the end of this
section.

To specify the model completely, we will assume 
\begin{align}
r_s(v) &= \begin{cases}k'\left(v_l-v\right)^{1/3} &\text{for } v\le v_l
\\
0 &\text{for } v> v_l
\end{cases}
\>.
\label{eq:decay_hawk}
\end{align}
This time dependence arises from the relationship for Hawking
radiation emitted by a macroscopic black hole and as seen by a distant
observer. The temperature of a black hole is inversely
proportional to its mass\cite{hawking75,opatrny12}
\begin{align}
  T_{\text{BH}}(M) &=\frac{\hbar c^3}{8 \pi G k_B M}\>,
  \label{eq:hawking_temp}
\end{align}
where Planck's and Boltzmann's constants
appear in standard notations. The thermal radiation of a black body at
this temperature is proportional to $T_{\text{BH}}^4$ but also to the
surface area of the black hole, which goes as
$r_s^2 \propto M^2 \propto T_{\text{BH}}^{-2}$, so the total power
output is proportional to $T_{\text{BH}}^{2}\propto M^{-2}$,
\begin{align}
P_{\text{BH}}(M) &=\frac{\hbar c^6}{15360 \pi G^2} M^{-2}\>,
\end{align}
from which we obtain a differential equation for the mass
\begin{align}
\abl{M}{v} &=-\frac{\hbar c^4}{15360 \pi G^2 M^{2}}\>,
\end{align}
which is solved by $M^3(v) = \tilde{k} \left(v_l-v\right)$, where 
\begin{align}
  v_l= 5120 M_0^3 \frac{\pi G^2}{\hbar c^4}
  \label{eq:lifetime_BH}
\end{align}
is the lifetime of the black hole, $M_0$ is its initial mass, and
$\tilde{k}= M_0^3/v_l$. The quantitiy $k'$ is then just
$2G\tilde{k}^{1/3}/c^2$ (so that $r_s(0) = 2 G M_0/c^2$). In
Ref.~\onlinecite{hiscock81a}, it is argued that the dependency
\eqref{eq:decay_hawk}, based on a fixed-background calculation, cannot
hold down to mass zero, as this would lead to an infinite flux of
radiated particles. At least near the end of evaporation, the
functional law must then be modified, an effect that we have studied
but which has little impact on the results given here, so we skip its
discussion.

The equations of motion for a test particle
falling freely in the Vaidya spacetime
may be obtained from the Lagrangian
\begin{align}
  \mathcal{L}&=\left(1-\frac{r_s(v)}{r}\right) c^2 \dot{v}^2
               - 2 \,c \,\dot{v} \dot{r}-r^2\left(\dot\vartheta^2+\sin^2\vartheta\dot\varphi^2\right)\>,
               \label{eq:lagrangian_geodes}
\end{align}
in which a dot denotes a derivative with respect to the proper time
$\tau$ of the particle. The Euler-Lagrange
equations read
\begin{align}
\abl{}{\tau}\pabl{\mathcal{L}}{\dot{q}}-\pabl{\mathcal{L}}{q}=0\>.
\end{align}
Writing them out for $q=\varphi$ and $q=\vartheta$ we find that they
are solved by $\varphi=\text{const.}$ and $\vartheta=\text{const.}$
Therefore, purely radial geodesics exist and to find these, we may
consider an effective Lagrangian, obtained from
Eq.~\eqref{eq:lagrangian_geodes} by dropping the last two terms.  The
equations for the two other coordinates then are
$\left(\ds r_{s,v}\equiv \D r_s/\D v=r'_{s}(v)\right)$:
\begin{align}
  \left(1-\frac{r_s}{r}\right) c\ddot{v} +\frac{r_s}{r^2} c \dot{v}\,\dot{r}
  - \ddot{r}-\frac{r_{s,v}}{2r} \,c \dot{v}^2&=0\>,
\label{eq:ddot_rv}\\
  \ddot{v} + \frac{r_{s}}{2r^2} \,c \dot{v}^2 &=0\>,
\label{eq:ddot_v}
\end{align}
and the definition of the effective Lagrangian will be useful for
simplifications of the equations of motion:
\begin{align}
  \left(\abl{s}{\tau}\right)^2&= c^2
                                = \left(1-\frac{r_s(v)}{r}\right) c^2 \dot{v}^2
                                - 2 \,c \,\dot{v} \dot{r}\>.
               \label{eq:lagrangian_effect}
\end{align}
This expresses that the four-speed of a massive particle is equal to
$c$. First, we use \eqref{eq:ddot_v} in \eqref{eq:ddot_rv} to
eliminate $\ddot{v}$ and then we replace the newly arisen term
$\left(1-\frac{r_s(v)}{r}\right) c^2 \dot{v}^2$ with the help of
\eqref{eq:lagrangian_effect} (by $c^2+2 \,c \,\dot{v} \dot{r}$), which
leads to a cancellation of all mixed terms containing the product
$\dot{v} \dot{r}$. The result is
\begin{align}
  \ddot{r} = -\frac{r_s(v)}{2r^2}\,c^2-\frac{r'_s(v)}{2r} \,c \dot{v}^2 \>.
  \label{eq:ddot_r}
\end{align}
Equations \eqref{eq:ddot_r} and \eqref{eq:ddot_v}, together with the
side condition \eqref {eq:lagrangian_effect} on the four-velocity
constitute the equations to be solved for radial fall towards the
evaporating black hole. They are equivalent to a set of three
first-order equations. In fact, it can be shown that
Eq.~\eqref{eq:ddot_v} is a consequence of Eqs.~\eqref{eq:ddot_r} and
\eqref{eq:lagrangian_effect} (at least for $r$ values greater than
$r_s(v)$). A unique solution requires initial conditions for $r$
(usually $>r_s$), $\dot r$ and $v$.

In the limit $r_s=\text{const.}$, i.e., for an eternal black hole, the
second term on the r.h.s.~of Eq.~\eqref{eq:ddot_r} is zero and
inserting the definition of the Schwarzschild radius, we obtain
$\ddot{r} = -GM/r^2$, which has the same form as Newton's law for the
free fall of a particle in the gravitational field of a point mass (or
a spherically symmetric mass distribution with radius smaller than
$r$). Since the time derivatives are with respect to proper time
rather than absolute Newtonian time, the dynamics will look Newtonian
only at sufficiently small velocities, as long as the difference
between the rate of proper time of the particle and that of an
external observer does not become visible. 

If we now take a temporal variation of the mass term $r_s$ into
account, then we have both contributions, the first being a
generalized Newtonian law with time-dependent mass term
($r_s(v)\leadsto M(v)$) and the standard $1/r^2$ dependence, whereas
the second term contains the time derivative of the mass and decays as
$1/r$. For diminishing mass, the second term is positive, so it is a
repulsive contribution to the acceleration of the test
particle. 

In a Newtonian universe, we would expect only the first term to be
present (with the advanced time in the argument of $r_s$ replaced by
absolute time). This should be true even when the mass varies with
time, because Newtonian gravity spreads instantaneously. Hence, what
we see in general relativity instead is an effect of gravity traveling at
a finite velocity.

Let us estimate the relative sizes of the second and first terms. For $v<v_l$, the
ratio between their magnitudes is
\begin{align}
  Q = \frac{\abs{r_s'(v)} r \dot{v}^2}{r_s(v) c} = \frac{r \dot{v}^2}{3 c(v_l-v)}
  \approx \frac{r \dot{v}^2}{3 cv_l}\>.
\end{align}
The last approximation is valid for black holes having a mass bigger
than that of the sun and up to times $v$ well exceeding the current
lifetime of the universe ($13.8\times 10^{9}\>\text{y}$), because we have 
$v_l > 2.1\times 10^{67}\>\text{y}=6.6 \times 10^{74}\>\text{s}$.
From Eq.~\eqref{eq:lagrangian_effect}, we obtain the estimates
$\dot{v}<1/\sqrt{1-\frac{r_s}{r}}$ for negative $\dot r$ and
$\dot v< (1+\sqrt 2)/(1-\frac{r_s}{r})$ for positive $\dot r<
c$. Hence, if we require the radius $r$ to be larger than $1.01 r_s$,
we have $\dot{v}^2 < 6/(1-\frac{r_s}{r})^2 < 10^5$. Setting an
upper limit for the radial coordinate by requiring
$r<10^{11}\>\text{ly}\approx 10^{27}\>\text{m}$, which is larger than
the particle horizon of the currently favored model of the universe,
we find that $Q<1.7\times 10^{-52}$ in the specified temporal and
spatial range.

On time scales of the current lifetime of the universe, $r_s$ is
constant for non-rotating astrophysical black holes
($M\ge M_{\astrosun}$, the mass of the sun), as long as nothing falls
on them.  The reason is that the power output of their Hawking
radiation is extremely small (below $10^{-28}\>\text{W}$), leading to
a negligible amount of radiated energy in comparison with their mass
energy for at least $10^{50}\>\text{y}$. Obviously, such a black hole
must be very well described by the Schwarzschild geometry, except
possibly (very) close to the horizon. Our estimate shows that the
ingoing Vaidya metric is, on the radial and time scales discussed,
indistinguishable from the Schwarzschild metric (written in
EF coordinates), as would be the outgoing Vaidya
metric with comparable parameters. Therefore, radial particle
trajectories in the spacetime of a non-rotating astrophysical black
hole will be described to an extremely good approximation by
Eq.~\eqref{eq:ddot_r} for $r>1.01 r_s$ and time scales of up to a few
billion years.

Closer to the horizon, the static approximation may be insufficient,
because the second term of Eq.~\eqref{eq:ddot_r} can become large. But
there, the ingoing Vaidya metric should be good, because it faithfully
reproduces the negative energy density near the horizon. According to
a fairly rigorous calculation in $1+1$ dimensions,\cite{davies76} the
sum of the outgoing and ingoing fluxes in the radial rest frame, which is proportional to
$T_{uu}+T_{vv}$, becomes zero at $r\approx 1.6 r_s$ (and is
negative for smaller $r$), so the Vaidya metric with its negative
energy density should be good up to about this radial coordinate, and
this should be true independent of the mass of the black hole.  This
is also consistent with Ref.~\onlinecite{dey19}. Hence, for black holes
of realistic size, approximating the metric via the ingoing Vaidya
metric will be viable for all interesting regions of
spacetime. Moreover, it has the advantage of analytic accessibility
that is lost, if instead we patch two or more metrics
together\cite{hiscock81b} in order to gain a tiny bit of accuracy at
larger $r$.

In order to render the time-dependent Schwarzschild radius visible in
our figures in the next section, we have to consider black holes with
much smaller masses than will be found in astrophysics. For these, we
expect the ingoing Vaidya metric to be a good approximation up to
$r=1.6 r_s$ and a decent approximation out to a few more Schwarzschild
radii, so answers obtained from it should be qualitatively
correct. Results will become qualitatively \emph{incorrect}, once the
second term of Eq.~\eqref{eq:ddot_r} becomes bigger than the first one
at \emph{large} $r$, due to the fact that the second term falls off
with a weaker power of $r$ than the first. The dominant second term
will then have the wrong sign, because it would be attractive for an
outgoing Vaidya metric.

\section{Numerical solution and analytic discussion}
\label{sec:num_sol_discuss}

To gain understanding about the dynamics in this metric, the equations
of motion should be integrated numerically for exemplary situations. They are
simple enough to make this a nice practical exercise.

In a number of cases, we solved the geodesic equations 
beyond the evaporation time, where the derivative of
$r_s(v)$ has a singularity $\propto~\!\!(v_l-v)^{-2/3}$.
This poses an obstacle to direct numerical integration.  Solvers for
ordinary differential equations with time step control will not
normally cross that point, in attempting to resolve the singularity.
Therefore, a numerical approach avoiding the appearance of diverging
quantities was developed. Its details will not be presented here, in
order to keep the paper concise.\footnote{Interested readers can find them in
an earlier version of this article, deposited on
arXiv: 2103.08340v2 [gr-qc].}

\begin{figure}[ht]
  \vspace{4mm}
  \includegraphics*[width=8.2cm]{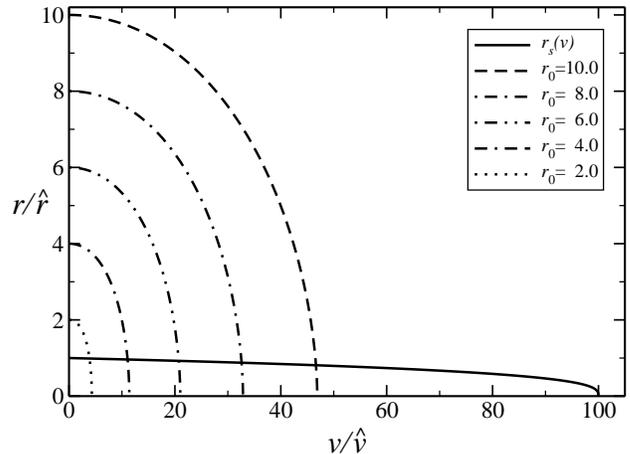}
  \caption{Trajectories of radially infalling particles in the
    Vai\-dya metric \eqref{eq:vaidya_metric}. $\hat{r}=r_s(0)$,
    $\hat{v}=r_s(0)/c$. The unit $\hat{r}$ has been left out in the
    legend for brevity, i.e., $r_0/\hat{r}$ has been renamed to $r_0$.}
  \label{fig:five_falls} 
  \vspace{2mm}
\end{figure}

In Fig.~\ref{fig:five_falls}, the trajectories of a particle released
from rest at different heights and approaching the Schwarzschild
radius well before evaporation are presented. The parameters are the
same as in Fig.~1 of Ref.~\onlinecite{kassner19} and the outcome is
similar. The particle falls into the black hole in all cases.

Our results are given in terms of nondimensional variables, with
$\hat{r}\equiv r_s(0)$ chosen as the length unit and the speed of
light set equal to $1$. Then the time unit is also fixed (at the value
$\hat{v}=r_s(0)/c$). By choosing a nondimensional value for the time
$v_l$ of evaporation, defined by $r_s(v_l)=0$, we therefore fix a
physical quantity, which is the initial mass of the black hole. To see
this, consider
\begin{align}
  \frac{v_l}{\hat{v}}=\frac{c v_l}{r_s(0)} = 2560 M_0^2 \frac{\pi G}{\hbar c} = 2560 \pi \frac{M_0^2}{m_p^2}\>,
\end{align}
where $m_p=\sqrt{\hbar c/G}=2.176 \times 10^{-8}\>$kg is the Planck
mass. For the values $v_l/\hat{v} =100$, $10$, and $5$ used in our
figures, we have $M_0 = 0.112\, m_p$, $0.0352\, m_p$, and
$0.0249\, m_p$, respectively, extremely small masses indeed. Black
holes with such a small mass cannot arise from direct gravitational
collapse. They might be the result of a density fluctuation in some
violent subatomic process, of a kind that could possibly have arisen
very shortly after the big bang. These black holes would, however,
have long since vanished. Yet, it is conceivable that some primordial
black holes starting out with masses around $1.8 \times 10^{11}\>$kg,
approaching the end of their life today, might have current
masses close to a Planck mass.

The reason for considering such unrealistic cases here is that for a
black hole of about one solar mass, the nondimensional $v_l$ would be
on the order of $10^{80}$, and the Schwarzschild radius would look
constant in our figures. 
To see a particle actually survive the evaporation of the black
hole, we must tune its starting distance and time so that it closes in
on the horizon position only near the point of complete evaporation
(or later).

\begin{figure}[ht]
  \vspace{3mm}
  \includegraphics*[width=8.2cm]{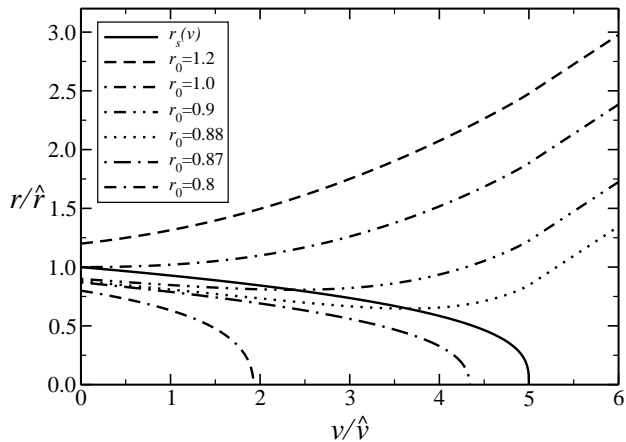}
  \caption{Light sent radially outward from positions near the initial
    Schwarzschild radius.}
  \label{fig:light_rays}
  \vspace{2mm}
\end{figure}

Before looking at this kind of behavior, let us briefly clarify
whether the Schwarzschild radius $r_s(v)$ is still an event horizon,
when $r_s$ is decreasing as a function of time. 
The equation of motion for an outgoing radial light ray follows from
Eq.~\eqref{eq:vaidya_metric} by setting $\D\Omega=0$, $\D s=0$, and $\D v\ne 0$:
\begin{align}
  \abl{r}{v} = \frac12 \,c \left(1-\frac{r_s(v)}{r}\right)\>.
\end{align}
This has been integrated in Fig.~\ref{fig:light_rays} for a few
initial values $r=r_0$ close to $r_s(0)$.

A light ray sent from the Schwarzschild radius $r_s(0)$ will obviously
escape, having zero coordinate velocity at $v=0$ and positive velocity
at any later time. Interestingly, some rays starting their journey
with $r<r_s(0)$ manage to cross the shrinking Schwarzschild radius,
which therefore is not a true event horizon anymore, but rather an
apparent horizon. An apparent horizon is a limiting trapped null
surface; i.e., outward-directed light rays emitted outside the
apparent horizon will move outward, while outward-directed light rays
emitted inside it will move inward.

Escaping light rays from inside start falling inward
but do so more slowly than the horizon shrinks, so they are passed
by the latter and afterwards move outward. Light sent from sufficiently far
inside this apparent horizon will, however, fall into the singularity,
so there still exists an event horizon, separating events from which
null infinity can never be reached from those that may send signals to
null infinity. That horizon, however, does not exist indefinitely; it is
gone after the time $v_l$. Moreover, if quantum mechanics leads to
avoidance of the singularity as is generally believed, possibly
replacing it with a fuzzy region of fluctuating spacetime that
contains quantum fields at high energy density, whatever was caught in
that high-energy density region would get out again (albeit transformed
into photons or other elementary particles), once the region dissolved, and
escape to future null or timelike infinity. But then there would be no event
horizon by definition.

\begin{figure}[ht]
  \vspace{3mm}
  \includegraphics*[width=8.2cm]{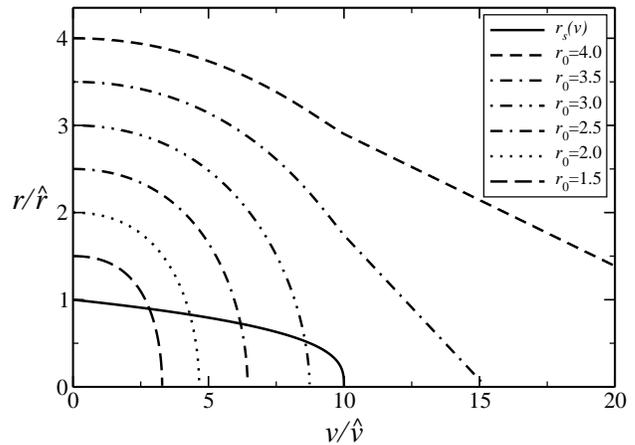}
  \caption{Trajectories missing and hitting the black hole.}
  \label{fig:six_falls}
  \vspace{2mm}
\end{figure}

Let us next consider a situation, where it is possible for a particle
to miss the singularity in the black hole. This is depicted in
Fig.~\ref{fig:six_falls}. Particles starting from a radius slightly
exceeding $3.2\, r_s(0)$ will not cross the apparent horizon. The
inward velocity of both the particle starting at $3.5\, r_s(0)$ and
that starting at $4\, r_s(0)$ decreases a little just before complete
evaporation, which is obviously due to the repulsive term in
\eqref{eq:ddot_r}.

The model can even produce overall repulsion after an initial phase of
inward falling.\cite{piesnack20} However, since for this to happen,
the second term of Eq.~\eqref{eq:ddot_r} must become larger than the
first one, due to $r$ getting large, the ingoing Vaidya metric will
not be a good model for an evaporating black hole anymore, so we will
not consider this case here.

Let us instead determine the range of $r_0$ values, for which the force on
the particle is repulsive already in the \emph{initial} phase of its
fall. Since we release it from rest, we may assume that
$\dot{r}\approx 0$. Using \eqref{eq:lagrangian_effect}, we then
express $\dot{v}^2$ by functions of $r$ and $v$ alone and find from Eq.~\eqref{eq:ddot_r}
\begin{align}
  \ddot{r} = -\frac{r_s(v)c^2}{2r^2}+\frac{r'_s(v)c}{2(r_s-r)}\>.
  \label{eq:ddot_r_rest}
\end{align}
Setting $\ddot{r}=0$, we obtain a quadratic equation for $r$ with the
roots\footnote{The approximation $r_s\ll r$ may be directly applied in
  \eqref{eq:ddot_r_rest} to obtain the approximate result for
  $r_{01}$. It also implies $r_{s,v}\ll c$, which has to be used in the
  equation for $r_{02}$, not satisfying $r_s\ll r_{02}$.  }
{\allowdisplaybreaks
  \begin{align}
  r_{01} &= -\frac{r_s c}{2 r_{s,v}} + \sqrt{\frac{r_s^2 c^2}{4 r_{s,v}^2} + \frac{r_s^2 c}{r_{s,v}}}
           \underset{r_s\ll r}{\approx} -\frac{r_s c}{r_{s,v}} = 3 c v_l\>,
            \label{eq:frst_repuls_rad}
  \\
  r_{02} &= -\frac{r_s c}{2 r_{s,v}} - \sqrt{\frac{r_s^2 c^2}{4 r_{s,v}^2} + \frac{r_s^2 c}{r_{s,v}}}
           \underset{r_{s,v}\ll c}{\approx} r_s\left(1+\frac{r_s}{3 c v_l}\right)\>. 
           \label{eq:sec_repuls_rad}
  \end{align}
}%
Repulsion dominates for $r_0>r_{01}$ and $r_0<r_{02}$. The first
result is irrelevant for stellar-mass evaporating black holes, because
it is exorbitantly large, bigger than $6.3\times10^{67}\>$ly, well
beyond $10^{50}$ times the particle horizon of the universe.

On the other hand, the second result tells us that there is repulsion
in the ingoing Vaidya metric with decreasing mass slightly outside,
but near, the Schwarzschild radius.  And this close to the horizon,
the ingoing Vaidya metric can be safely assumed to yield a decent
description!

\begin{figure}[ht]
  \vspace{3mm}
  \includegraphics*[width=8.2cm]{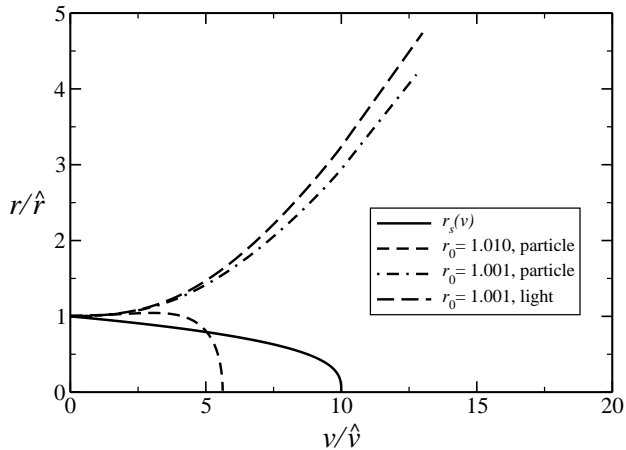}
  \caption{Trajectories of particles released from rest at
    $r_0=1.01\, r_s(0)$ and $r_0=1.001\, r_s(0)$. The second particle
    escapes to infinity. The path of a light ray starting from the
    same $r_0$ as the second particle is given for comparison.}
  \label{fig:comp_part_light}
  \vspace{2mm}
\end{figure}

We first verify the validity of the analytic prediction by
numerical simulation. Fig.~\ref{fig:comp_part_light} gives the
trajectories, for $v_l/\hat{v}=10$, of a particle starting at a
radius that exceeds $r_s(0)$ by just one percent and of one starting
only one thousandth of $r_s(0)$ above the apparent
horizon. Eq.~\eqref{eq:sec_repuls_rad} predicts initial repulsion for
$r_s<r_0<1.033\>r_s$ here.
 

The first particle starts moving away from the apparent horizon, being
repelled more strongly than attracted, but as its distance from
$r_s(v)$ increases, it is pulled back, turns around, and finally falls
into the singularity. However, the second particle, starting
significantly \emph{closer} to the apparent horizon, experiences such
a strong repulsion that it actually escapes from the black hole. In
order to gain some understanding, we have also plotted the trajectory
of an outgoing light ray starting together with the second
particle. What becomes immediately clear is that the two trajectories
are pretty close to each other, so the particle reaches a speed that
is close to the speed of light. It is then easy to form a meaningful
mental picture by remembering the river model of a black hole,
proposed by Hamilton and Lisle.\cite{hamilton08} According to this
idea, space about a black hole may be viewed as flowing inward across
the event horizon, where it reaches the speed of light. This explains
nicely why nothing can cross the event horizon from the inside out --
motion relative to space cannot exceed the speed of
light.\footnote{This is true for classical objects. For quantum
  tunneling, the speed of light is not an insurmountable barrier.}
With the Vaidya metric, inflowing space reaches the speed of light at
the time-dependent apparent horizon rather than at the event
horizon,\footnote{This holds for other metrics with a receding apparent
  horizon as well.\cite{kassner19}{}} but otherwise the situation is
similar. Positioning a particle at rest slightly above the apparent
horizon actually means giving it a very high outward speed with
respect to the inflowing space.  The closer to the apparent horizon
this is done, the closer the outward speed is to the speed of light. A
particle extremely close to the apparent horizon should therefore move
like a photon. Since outgoing light will always escape when emitted
outside the horizon, a massive particle close enough to it will do the same.

Note that it is \emph{impossible} to start a particle from rest
\emph{inside} the apparent horizon, as its velocity with respect to the
inflowing space would have to exceed $c$.
However, it is possible to give a particle slightly inside, and close
to, the apparent horizon an initial condition corresponding to a
sufficiently small inward velocity so that the shrinking horizon will
catch up with it, allowing it to escape to infinity. Indeed, we find
in simulations (not shown) that a particle which is started slightly
inside the initial Schwarzschild radius with a coordinate velocity not
significantly below $u_0=-c\sqrt{r_{s}(0)/r_0-1}$
manages to cross the apparent horizon and to escape to future timelike
infinity.

We expect this result to be robust with regard to a modification of
the metric outside of a neighborhood of $r_s$. Any outside metric
patched to the ingoing Vaidya metric must not add an additional
horizon. Light that reaches the outer metric will therefore escape to
future infinity. A particle accelerated to relativistic velocities
by the inner metric will then generally also escape to infinity on
entering the outer metric.

\section{Conclusions}
\label{sec:conclus}

The Vaidya metric is an appealing tool to present certain effects of
non-stationary gravitational fields in class\-room, having
applications to black-hole evaporation.  Simple analytical results
demonstrate interesting physical effects, in particular gravitational
repulsion in the presence of a shrinking apparent horizon. The
numerical studies presented here are not too advanced, and can be
assigned to students as exercises.

Use of the ingoing Vaidya metric with a decreasing mass term is
motivated by its utility in modeling a spacetime geometry with a
shrinking future (apparent) horizon.  We have argued that this kind
of description is reasonable near and outside the apparent horizon,
because negative energy density is expected to arise there due to
quantum effects,\cite{hawking75,davies76} and the Vaidya metric
considered describes an inflow of negative energy carried by null
dust. If we take Hawking's view that this is the same as an outflow of
positive-energy particles along past-directed geodesics, then we can
say that what the ingoing Vaidya metric fails to account for is the
final scattering of these quanta off the gravitational field that
makes their path future-directed again. This would certainly be bad,
if the purpose was to describe these particles, i.e., Hawking
radiation itself. However, we were mainly interested in physical
effects of the modifications of the metric brought about by Hawking
radiation. For most of the time a black hole exists, Hawking
radiation is a small effect. The metric at some distance from the horizon will
essentially be the Schwarzschild metric; i.e., $r_s$ changes so slowly
that the second term on the right-hand-side of Eq.~\eqref{eq:ddot_r}
is negligible and it also does not really matter whether the argument
of the first term is the advanced time $v$, the retarded time $u$ or
even the Schwarzschild time $t$.

This is no longer true near the end of the black hole's life, because
then Hawking radiation becomes a strong effect and even may dominate
the metric. We must then model it more accurately in the whole
spacetime. It is therefore concluded that results (not discussed here)
such as net repulsion after full evaporation\cite{piesnack20} are an
artifact of the ingoing Vaidya metric in most cases. This effect is
not likely to occur for a black hole at the end of its life.

A more sophisticated approach, properly reproducing the
(future-directed) outflow of positive energy far from the horizon
would be to keep the ingoing Vaidya metric for radii up to a small
multiple of $r_s$ and use an outgoing Vaidya metric for larger
$r$.\cite{hiscock81b} There would be matching conditions for the
places where the two metrics meet. Realizing this approach might be a
bit too involved for the classroom, and it would not provide any
quantitative gains for the description of realistic astrophysical
black holes.

The properties of evaporating black holes obtained within our model
and discussed here should be robust against model variations aiming at
a generally more quantitative description.  Two of these are pretty
intuitive. The first is that, for particles falling towards the black
hole sufficiently long before evaporation, there is no problem
crossing the horizon. Hawking radiation does not prevent them from
falling in. This would be expected from a comparison of the time
scales on which Hawking radiation becomes relevant with those for an
infaller to cross the horizon, if a time coordinate is used that does 
not become singular at the horizon. Second, the nature of the horizon
changes. The event horizon of the stationary limit of the Vaidya
metric (which is the corresponding EF metric) turns
into an apparent horizon. That is, it remains a trapped null surface,
but light may eventually escape from it due to its shrinking in
time. Contrary to apparent horizons of growing black holes (the normal
case), this apparent horizon is \emph{outside} of the event horizon,
obviously because the standard condition of positive energy density
everywhere is violated.

Moreover, it is expected that a counterintuitive aspect found here
will prevail in a generally more quantitative description,
\emph{viz.}~the strong repulsion experienced by a particle released from rest
just above the horizon. Such a particle may escape to infinity and
this effect, predicted from the Vaidya metric (but also from a
time-dependent generalization of the GP metric\cite{kassner19}) is
likely to be a qualitatively correct result. Once the particle has
been accelerated to nearly the speed of light in a thin shell above
the horizon, it will also escape in a model, where the outer part of
the metric has been changed into something more realistic. Because
this outer metric will not contain a second horizon, any relativistic
particle should not have problems escaping, just as a light ray would.

Arguably, the effect is not so counterintuitive after all, because of
the presence of negative energy density. Repulsion in the
Reissner-Nordström metric is explicable because of negative pressure
(which is stress),\cite{gron86} exerted by the electric field. In the
equations of motion, an electrostatic potential term appears,
representing negative energy. This then suggests that it is the
negative energy density in both the Reissner-Nordström solution and
the Vaidya metric that is responsible for the appearance of
repulsion. A difference, however, is that the electrostatic energy
term in the Reissner-Nordström geodesic equations exceeds the mass
energy of the black hole inside the inner horizon. By contrast, a
volume integral of the negative energy density of the Vaidya metric,
constructed to represent an evaporating black hole, will be much
smaller than the remaining mass energy of the black hole for all times
except very briefly before complete evaporation. Moreover, the
repulsion will happen only for particles almost coordinate stationary
near the horizon. An infalling particle having already acquired a
substantial inward velocity at the horizon will not experience
repulsion ($\dot v$ remains finite at $r=r_s$ in
\eqref{eq:lagrangian_effect} for $\dot r< 0$). In the
Reissner-Nordström case, the repulsion is velocity independent. These
considerations suggest that the presence of negative energy density
alone does not explain the repulsion effect in the case considered
here, although it may be a necessary condition for radial repulsion.

On the other hand, there is no need to invoke energy considerations at
all to \emph{understand} repulsion the way we presented it in discussing
Fig.~\ref{fig:comp_part_light}. If the apparent horizon is receding,
no matter what the reason, then a coordinate stationary particle close
enough to it will have essentially the speed of outgoing light
relative to the inflowing space in the river model of a black
hole.\cite{hamilton08} Such a relativistic particle will be ejected
and is likely to escape from the black hole. In this view, the
repulsion is not due to a repelling force but rather an inertial
effect. By contrast, a particle that has a sufficiently large
inward velocity slightly outside the horizon is much slower than light
with respect to the inflowing space and thus will not be able to avoid
being captured.

Finally, the system considered here seems to be the simplest one
exhibiting radial gravitational repulsion outside a horizon. In
principle, our prediction of particles being ejected violently from
near the horizon could be tested in a sonic black-hole
analog,\cite{unruh81} set up with a receding horizon.

\end{document}